\documentclass[twocolumn,english,aps,prb,twocolum,superscriptaddress,bibnotes,amsmath,amssymb,floatfix]{revtex4-2}
\usepackage[colorlinks=true,citecolor=blue,linkcolor=magenta]{hyperref}

\usepackage{changes}
\usepackage{soul}
\usepackage[utf8]{inputenc}
\usepackage[english]{babel}
\usepackage{amsmath,amsfonts,amssymb}
\usepackage[T1]{fontenc}
\usepackage{url}

\usepackage{amsmath}
\usepackage{amsfonts}
\usepackage{amssymb}
\usepackage[version=3]{mhchem}
\DeclareUnicodeCharacter{2009}{ }

\usepackage{epstopdf}
\usepackage{graphicx}
\graphicspath{{./Figures/}}
\begin{document}
% \title{Three-dimensional integration enabling ultra-low-noise, isolator-free silicon photonic integrated circuits}
\title{Turnkey locking of quantum-dot lasers directly grown on Si}
% \title{Three-dimensional integration and isolation of ultra-low-noise lasers for silicon photonic integrated circuits}

\author{
Bozhang Dong$^{1,\ast}$,
Yating Wan$^{2, \ast, \dagger}$,
Weng W. Chow$^{3}$,
Chen Shang$^{1}$,
Artem Prokoshin$^{2}$,
Rosalyn Koscica$^{4}$,
Heming Wang$^{1}$, and
John E. Bowers$^{1,4,5,\dagger}$,\\
\textit{
$^1$Institute for Energy Efficiency, University of California, Santa Barbara, CA, USA\\
$^2$Integrated Photonics Laboratory, King Abdullah University of Science and Technology, Thuwal, Makkah Province, Saudi Arabia\\
$^3$Sandia National Laboratories, Albuquerque, NM, USA.\\
$^4$Materials Department, University of California, Santa Barbara, CA, USA\\
$^5$Department of Electrical and Computer Engineering, University of California, Santa Barbara, CA, USA\\}
$^\dagger$Email: bowers@ece.ucsb.edu and yating.wan@kaust.edu.sa\\
$^*$These authors contributed equally\\
}
% A fully referenced ~200 word summary paragraph; main text of 2,500 words and 4 modest display items (figures, tables) for a typical 6 page article and 4300 words and 5-6 modest display items for a typical 8 page article; as a guideline up to 50 references if needed and within the allocated page budget

\begin{abstract}
Ultra-low-noise laser sources are crucial for a variety of applications, including microwave synthesizers, optical gyroscopes, and the manipulation of quantum systems. Silicon photonics has emerged as a promising solution for high-coherence applications due to its ability to reduce system size, weight, power consumption, and cost (SWaP-C). Semiconductor lasers based on self-injection locking (SIL) have reached fiber laser coherence, but typically require a high-Q external cavity to suppress coherence collapse through frequency-selective feedback. Lasers based on external-cavity locking (ECL) are a low-cost and turnkey operation option, but their coherence is generally inferior to SIL lasers. In this work, we demonstrate quantum-dot (QD) lasers grown directly on Si that achieve SIL laser coherence under turnkey ECL. The high-performance QD laser offers a scalable and low-cost heteroepitaxial integration platform. Moreover, the QD laser's chaos-free nature enables a 16 Hz Lorentzian linewidth under ECL using a low-Q external cavity, and improves the frequency noise by an additional order of magnitude compared to conventional quantum-well lasers.

\end{abstract}
\maketitle

Narrow-linewidth and frequency-stable lasers have opened up new opportunities for optical sensing and signal generation applications, including optical frequency and microwave synthesis \cite{spencer2018optical,Marpaung:19}, laser gyroscopes \cite{gundavarapu2019sub,lai2020earth}, light detection and ranging (LIDAR) systems \cite{trocha2018ultrafast,suh2018soliton,lihachev2022low}, spectroscopy \cite{suh2016microresonator}, and coherent optical communications \cite{olsson2018probabilistically}. In the quest for compact and mass-manufacturable coherent systems, photonic integrated circuits (PICs) have emerged as the most compelling solution. Semiconductor lasers with high-performance characteristics are integral components in the operation of PICs. Nevertheless, the linewidth of conventional semiconductor lasers ranges from hundreds of kilohertz to a few megahertz, which is more than seven orders of magnitude higher than the state-of-art of bulk fiber lasers \cite{matei20171}. The significant performance gap has prevented their widespread utilization in the aforementioned applications. Self-injection locking (SIL) and Pound-Drever-Hall (PDH) locking technologies \cite{dahmani1987frequency,hollberg1988modulatable,laurent1989frequency,Jin2021hertz,lihachev2022platicon,guo2022chip} have dramatically reduced the frequency noise and laser linewidth of semiconductor lasers. SIL, a passive locking technique, employs resonant optical feedback from a high-Q external optical element to suppress the coherence collapse of the pump laser and enabled the reduction of laser linewidth below that of a fiber laser \cite{li2021reaching}. However, The SIL operation also requires that the pump laser frequency coincides with the external cavity resonance, introducing  complexity and causing power penalties. These limitations need to be addressed to make this solution accessible to the industry.

%%%%%%%%%%%%%%%%%%%%%%%%%%%%%%%%%%%%%%%%%%%%%%%%%%%%%%%%%%%%%%%%%%%%%%
\begin{figure*}[t!]
\centering
\includegraphics[width=0.9\linewidth]{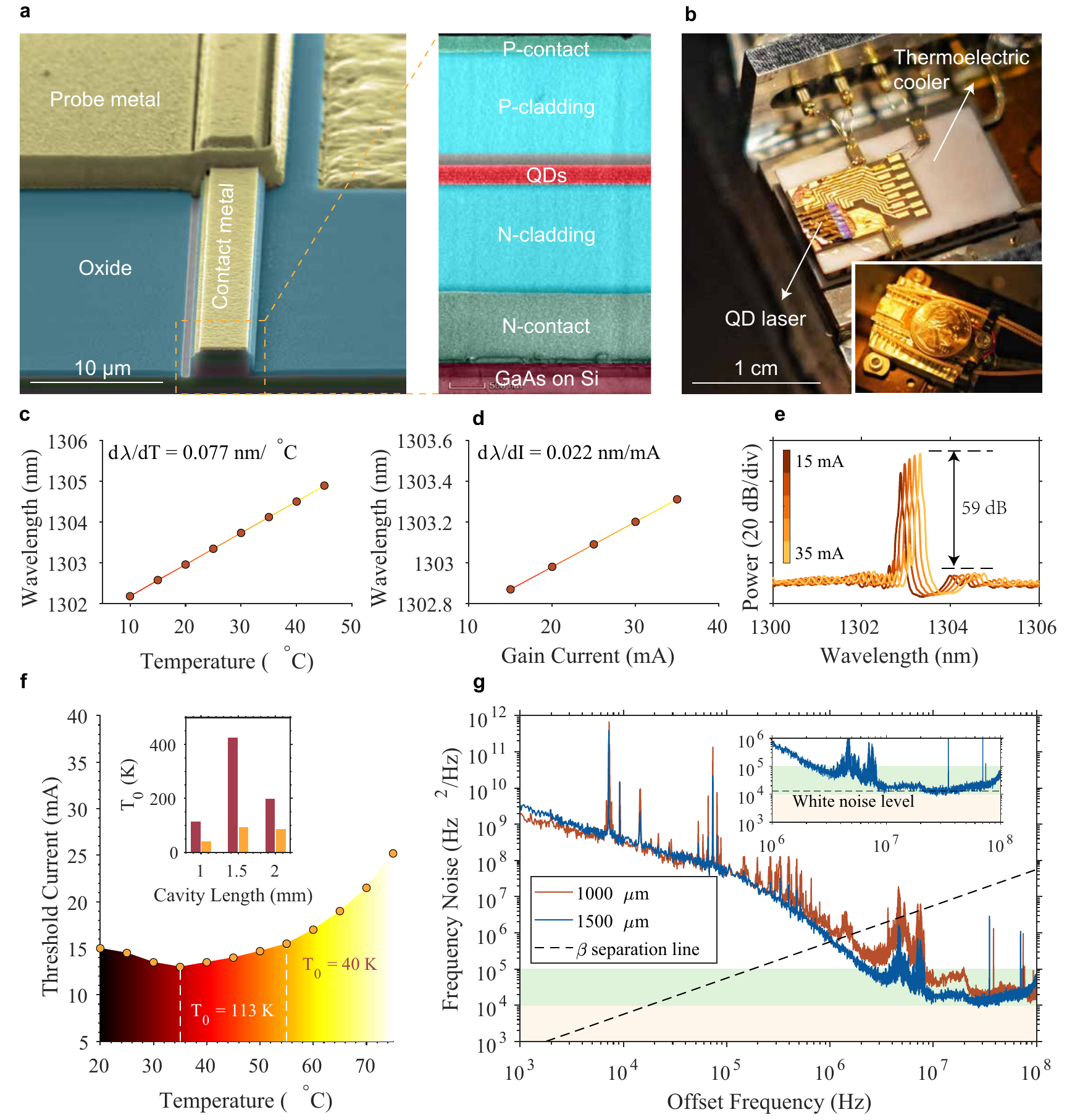}
\caption{\textbf{QD laser design and characterizations.}
\textbf{a}. Cross-sectional view of the axis (001) bright-field STEM image.
\textbf{b}. Images of the QD laser in a compact butterfly package.
\textbf{c}. DFB wavelength as a function of the operating temperature in a device with 3$\times$1000 $\mu m^2$ cavity.
\textbf{e}. Evolution of the optical spectra by increasing the injection current from the threshold current in a device with 3$\times$1000 $\mu m^2$ cavity. The optimum side-mode-suppression ratio (SMSR) is 59 dB. The threshold current at room temperature (20$^{\circ}$C) is 15 mA.
\textbf{f}. Threshold current as a function of the operating temperature. Inset: characteristic temperature T$_0$, in the ranges (35 - 55$^{\circ}$C and 55 - 80$^{\circ}$C) with different cavity lengths. T$_0$ = 423 K within 35 - 55$^{\circ}$C for the device with 1.5 mm cavity length.
\textbf{g}. Frequency noise spectra of the 1 mm device (burgundy) and another 1.5 mm-long QD laser (navy blue). The output power is fixed at 0 dBm. Inset: zoom in on the white noise level. The free-running QD laser yields a Lorentizian linewidth of 41 kHz.
}
\label{Fig:1}
\end{figure*}
%%%%%%%%%%%%%%%%%%%%%%%%%%%%%%%%%%%%%%%%%%%%%%%%%%%%%%%%%%%%%%%%%%%%%%

In this work, we introduce advanced quantum-dot (QD) lasers as a compelling solution for high-coherence turnkey external cavity locking (ECL) technology. Semiconductor QDs, due to the atom-like density of states, exhibit much lower linewidth enhancement factor (LEF), and thus much lower frequency noise and linewidth \cite{dong2021dynamic}. Furthermore, QD lasers offer higher immunity to the non-radiative defects between the III-Vs and Si, making them suitable for monolithic integration on Si substrates, particularly with high-Q SiN microresonators \cite{liu2015quantum,shang2022electrically}. This integration approach is more cost-effective compared to the heterogeneous integration approach, eliminating the need for expensive III-V wafers \cite{shang2021perspectives,zhou2023prospects}. The chaos-free feature of QDs reduces the dependence on a high-Q microresonator for linewidth locking, distinguishing QD lasers from conventional QW lasers. By leveraging the high-performance QD laser, we achieved a 16 Hz Lorentzian linewidth, the smallest number ever achieved for on-chip QD lasers. Four orders of magnitude reduction in frequency noise have been achieved using a low-Q external cavity without a frequency filter. On the contrary, QW laser, which is susceptible to coherence collapse under weak external feedback, requires a much larger external cavity Q-factor to maintain coherence. Leveraging the high feedback insensitivity of QD lasers, we improve the coherent feedback strength in the ECL operation, which allows for an extra one order of magnitude reduction in the frequency noise compared to its QW counterpart. Notably, our low-Q approach allows for turnkey locking without any power penalty, distinguishing it from the SIL approach that relies on a high-Q external cavity. Additionally, this approach offers greater flexibility for material platforms in which fabricating an ultra-high-Q resonator is challenging. This work effectively addresses the limitations of SIL technology and significantly enhances the SWaP-C metrics of high-coherence PICs. In addition, our theoretical investigations indicate that the full potential of the QD laser has yet to be realized. By integrating them with CMOS-ready high-Q microresonators, it becomes feasible to further reduce the linewidth to the sub-hertz level. These advancements position QD lasers as a compelling candidate for various high-coherence applications, including optical frequency synthesizers, dual-comb spectroscopy, and ultra-high-capacity optical transceivers. 

\medskip
\noindent \textbf{Results} 

\medskip
\noindent \textbf{Epitaxial growth of QD DFB lasers on Si}

The QD-on-Si laser material was grown in a Veeco Gen-II solid source molecular beam epitaxy (MBE) chamber on an anti-phase-domain free GaP/Si on-axis template. After completing the defect-reducing buffer layers and the QD active region \cite{shang2021high}, the sample is removed from the chamber for grating patterning. A uniform first-order DFB grating, with a duty cycle of 50\%, an etch depth of 50 nm, and a period of 197 nm, was dry etched into the top GaAs layer by inductively coupled plasma with a Cl$_2$/N$_2$ based chemistry. An electron beam lithography patterned SiO$_2$ hard mask was used to transfer the pattern. Before the sample was loaded back into the chamber for regrowth of the top cladding and the contact layers, a rigorous surface treatment combining solvent clean, O$_2$ plasma ash, and HF dip was performed. Thereafter, an in-situ atomic H clean for 30 min at 450$^{\circ}$C was conducted to remove any native oxide. The second MBE growth then proceeded with a 2 nm p-GaAs nucleation layer and the rest of a standard GaAs/Al$_{x}$Ga$_{1-x}$As graded index separate confinement heterostructure. The as-grown material was then processed into deeply etched waveguides with ridge widths ranging from 1.7 to 3 $\mu m$. The STEM image and the epi-structure of the QD laser are shown in Fig. \ref{Fig:1}a. The QD chipset is then mounted into a butterfly package (Fig. \ref{Fig:1}b) to facilitate measurements and enable portability.

\medskip
\noindent \textbf{High-performance integrated QD lasers}

Owing to the quantized density-of-state, the carriers are tightly confined in the dots leading to enhanced injection efficiency \cite{wan2021high}. From a series of devices measured, a maximum power of 11.7 mW is obtained from a 5$\times$1500 $\mu m^2$ cavity, twice the previously reported value \cite{wan20201}. The lowest threshold current of 10 mA is achieved from a 2.5$\times$800 $\mu m^2$ cavity. The lowest threshold current density of 400 A/cm$^2$ is obtained from a 2$\times$1500 $\mu m^2$ cavity, which corresponds to 80 A/cm$^2$ per QD layer. At room temperature (20 $^{\circ}$C), the QD laser exhibits a stable single-mode emission in the presence of a side-mode-suppression-ratio above 50 dB above the threshold (Fig. \ref{Fig:1}e). An SMSR of 59 dB is achieved when the laser operates at 35 mA. The current coefficient of the Bragg wavelength is as low as 0.022 nm/mA (Fig. \ref{Fig:1}d). It should be noted that the threshold current of a DFB laser is not only determined by the gain medium but also dependent on the optical mismatch between the laser gain peak and the Bragg wavelength \cite{dong2021dynamicdfb}. Such an optical mismatch can be detuned by varying the operating temperature due to the different temperature coefficients of the gain and the DFB wavelengths. Figure \ref{Fig:1}c depicts the evolution of the Bragg wavelength caused by the temperature variation. The extracted temperature coefficient of 0.077 nm/$^{\circ}$C is much lower than the temperature coefficient of the optical gain peak. As a result of the reduction in the optical mismatch, the threshold current is reduced from 15 to 13 mA as the temperature increases from 20 to 35 $^{\circ}$C (Fig. \ref{Fig:1}f). On the other hand, it should be noted that the requirement of laser cooling is a major challenge for PICs due to laser degradation at elevated temperatures. Such an inconvenience comes from both the increased nonradiative carrier recombination and the reduced gain due to the wider spreading of the Fermi distribution of carriers when they are heated up. The three-dimensional carrier confinement in III-V QDs equips these nanoparticles with superior insensitivity to the growth defects for the significantly reduced carrier diffusion length compared to QWs \cite{selvidge2019non}. The large energy separation between the ground and the excited state from the tight carrier confinement also enables better thermal stability \cite{shang2021high}. A QD laser with a 1 mm-long cavity yields a characteristic temperature $T_0$ of 113 K in the range of 35 to 55 $^{\circ}$C, and $T_0$ of 40 K in the range of 55 to 80 $^{\circ}$C. A moderate cavity length of 1.5 mm allows for improving the $T_0$ to 423 K and 92 K in the range of 35 to 55 $^{\circ}$C and 55 to 80 $^{\circ}$C, respectively (Fig. \ref{Fig:1}f). It should be noted that the $T_0$ extracted in the elevated temperature range is underestimated due to the additional increase in the threshold current induced by the aforementioned optical mismatch. In addition to the lasing performance, the QD laser is also beneficial to coherent applications owing to its near-zero LEF \cite{duan2018semiconductor}. It is worth stressing that the QD laser does not suffer from any degradation in the LEF when it is directly grown on Si. In the case of a single-frequency DFB laser, it is crucial to reduce the optical mismatch to take advantage of the intrinsic LEF of the active region. Figure \ref{Fig:1}g depicts the frequency noise spectra of the QD DFB laser with a cavity length of 1 mm (burgundy) and 1.5 mm (navy blue). The fiber-coupled power is fixed to 0 dBm and both devices operate at room temperature where the optical mismatch approaches zero. Both devices exhibit remarkable frequency noise in the presence of the white noise level below 1.8$\times 10^4$ Hz$^2$/Hz. A slightly better linewidth is observed in the 1.5 mm-long QD laser, where the white noise level is as low as 1.3$\times 10^4$ Hz$^2$/Hz, yielding a Lorentzian linewidth of 41 kHz. 

%%%%%%%%%%%%%%%%%%%%%%%%%%%%%%%%%%%%%%%%%%%%%%%%%%%%%%%%%%%%%%%%%%%%%%
\begin{figure*}[t!]
\centering
\includegraphics[width=\linewidth]{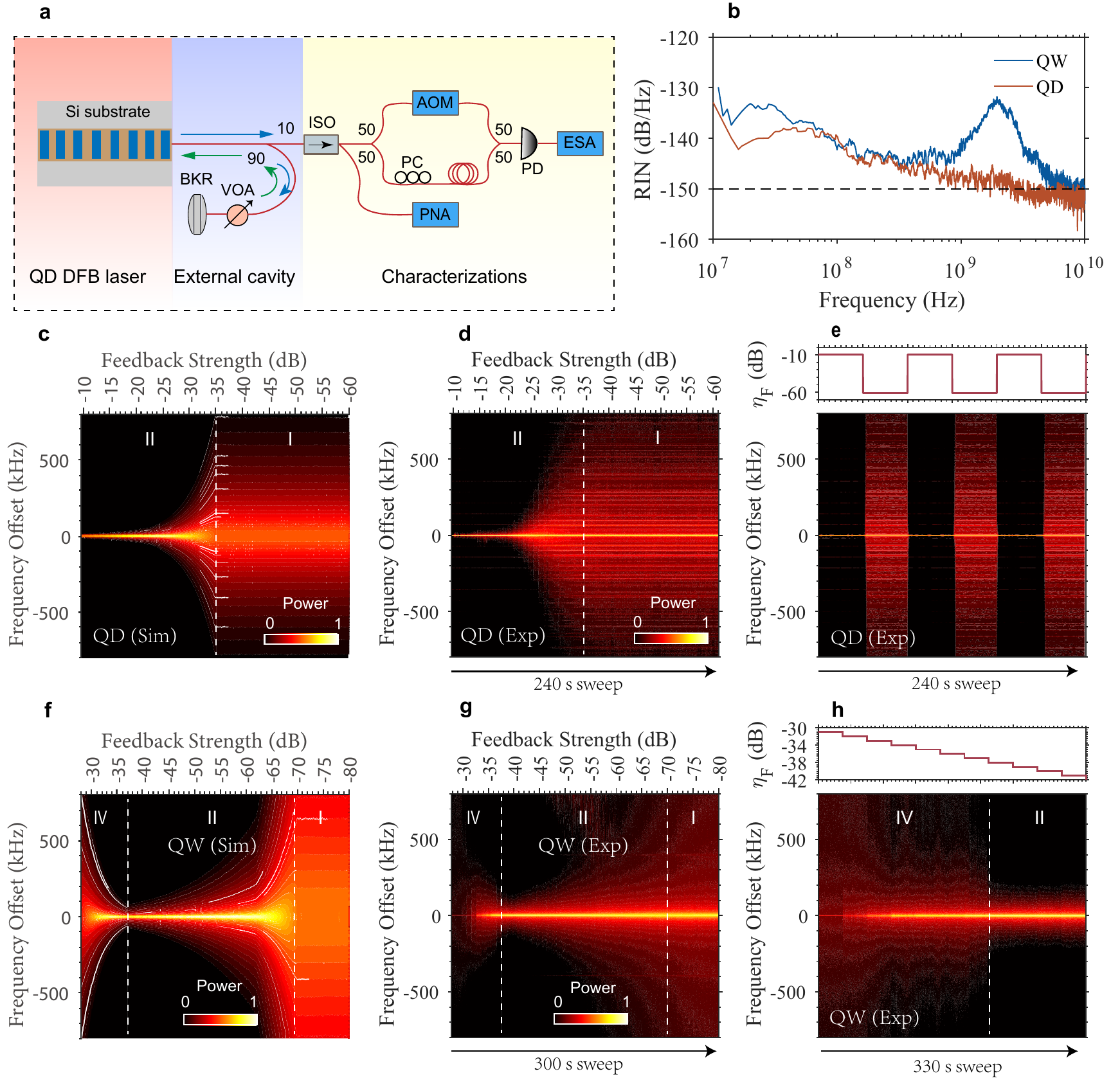}
\caption{\textbf{Laser dynamics under external optical feedback.}
\textbf{a}. Experimental setup for the external cavity locking, including an off-chip external cavity and a self-heterodyne setup for linewidth characterizations. VOA, variable optical attenuator; BKR, back reflector; ISO, optical isolator; PC, polarization controller; AOM, acoustic-optics modulator; PD, photodiode; ESA, electrical spectrum analyzer; PNA, phase noise analyzer.
\textbf{b}. Relative intensity noise of the free-running QD (burgundy) and the QW (navy blue) laser. The output power of both lasers is fixed at 0 dBm.
\textbf{c}. Theoretically calculated mapping of the spectral linewidth as a function of the feedback strength of the QD laser.
\textbf{d}. Experimentally measured mapping of the spectral linewidth as a function of the feedback strength of the QD laser.
\textbf{e}. Turnkey operation of the ECL QD laser under a modulation of the feedback strength between -60 and -9.6 dB.
\textbf{f}. Theoretically calculated mapping of the spectral linewidth as a function of the feedback strength of the QW laser.
\textbf{g}. Experimentally measured mapping of the spectral linewidth as a function of the feedback strength of the QW laser.
\textbf{h}. Transition from regime II to regime IV of the QW laser.
Regime I, stable operation regime; Regime II, external cavity locking regime; Regime IV, coherence collapse regime.
}
\label{Fig:3}
\end{figure*}
%%%%%%%%%%%%%%%%%%%%%%%%%%%%%%%%%%%%%%%%%%%%%%%%%%%%%%%%%%%%%%%%%%%%%%

\medskip
\noindent \textbf{Chaos-free operation under external optical feedback}

Laser dynamics under external optical feedback (EOF) have been extensively studied during the past four decades. The stability and instabilities of semiconductor lasers under EOF have opened up various applications. The coherent feedback allows for an increase in the intracavity photon density, which is beneficial to significantly reduce the laser linewidth. Nevertheless, a strong EOF would possibly lead to a severe coherence collapse. To take advantage of the narrow linewidth under EOF, it is crucial to suppress the coherence collapse of the laser. Compared to the conventional QW laser, the chaos-free QD laser opens up new opportunities for laser linewidth locking. 
Figure \ref{Fig:3}d depicts the laser linewidth as a function of the on-chip feedback strength for a wavelength-detuned QD laser while fixing its output power at 0 dBm. The experimental setup for the EOF and the linewidth characterization is shown in Fig. \ref{Fig:3}a. The QD laser remains in regime I in which the linewidth is unaffected when the feedback strength is below -35 dB. As the on-chip feedback strength increases from -35 to -9.6 dB, which is achievable by taking advantage of the low-loss SiN waveguide, a significant linewidth reduction takes place and the QD laser enters into regime II. Upon a periodic modulation of the feedback strength between -60 and -9.6 dB, the QD laser exhibits a turnkey transition from the unlocked state to the locked state and vice versa, as shown in Fig. \ref{Fig:3}e. On the contrary, the QW laser is much more sensitive to the EOF than the QD laser. Figure \ref{Fig:3}g depicts the evolution of laser linewidth due to external feedback for a commercial QW laser. The output power of the QW laser is fixed at 0 dBm for a fair comparison. Despite the ECL regime (regime II) when the feedback strength is in the range of -70 and -37 dB, the QW laser suffers from coherence collapse (regime IV) when the feedback strength is above -37 dB. A zoom-in of the transition from regime II to regime IV is shown in Fig. \ref{Fig:3}h. 

\begin{figure*}[t!]
\centering
\includegraphics[width=0.9\linewidth]{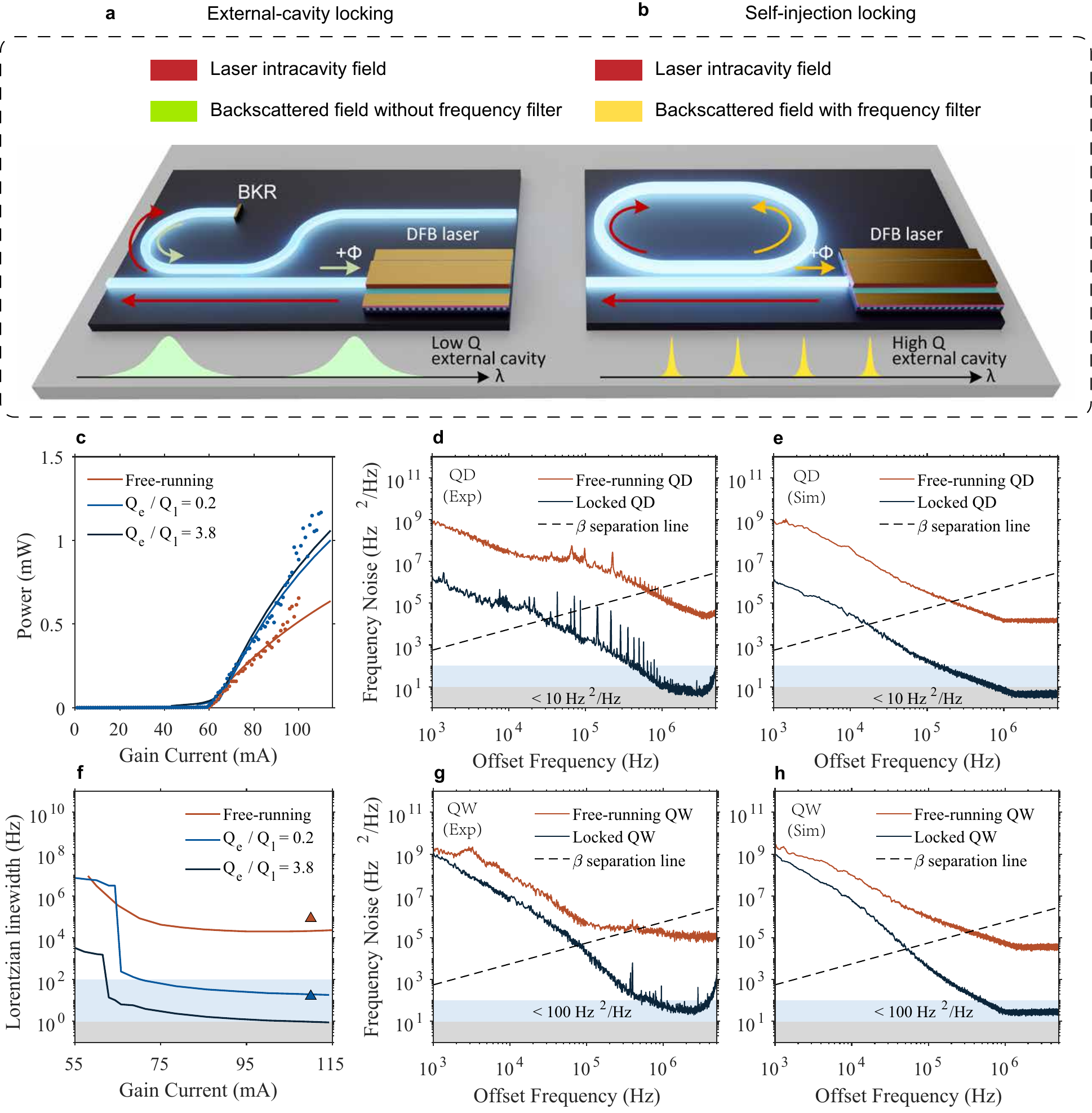}
\caption{\textbf{Turnkey external cavity locking operation.}
\textbf{a}. Operation of turnkey external cavity locking by using a low-Q external cavity. A back-reflector is used to reflect the laser emission (red arrow) back to the laser cavity. Laser operating in the feedback regime II allows the delayed laser field (green arrow) to go back to the laser cavity and interacts with the intracavity field for linewidth reduction. In this work, the external cavity is developed by using optical fiber and a fiber-based reflector.
\textbf{b}. Operation of self-injection locking by using a high-Q microresonator. A high-Q external cavity provides a narrow frequency filter for the pump laser, which enables frequency-selective external feedback. The pump laser light (red arrow) should be controlled to anchor to one of the external cavity resonances to generate a reflected field. The backscattered coherent light (yellow arrow) is reinjected into the laser cavity and interacts with the intracavity field for linewidth reduction. 
\textbf{c}. Calculated light-current characteristics (LI, solid curves) of the wavelength-detuned QD laser in free-running (burgundy) and in ECL operation. In both low-Q (navy blue) and high-Q (black) locking conditions, the QD laser doesn't suffer from a power penalty. The measured LIs are shown by the colored bullet point symbols.
\textbf{d}. Frequency noise spectra of the QD laser in free-running (burgundy) and in ECL (black) operation in the experiment.
\textbf{e}. Calculated frequency noise spectra of the QD laser in free-running (burgundy) and in ECL (black) operation.
\textbf{f}. Calculated Lorentzian linewidth as a function of injection current in free-running (burgundy) and in ECL operation. A low-Q locking condition (navy blue) in this work allows for a 16 Hz linewidth. A high-Q condition (black) can further reduce the laser linewidth to hertz level. Experimental results are shown by the colored triangle symbols.
\textbf{g}. Frequency noise spectra of the QW laser in free-running (burgundy) and in ECL (black) operation in the experiment.
\textbf{h}. Calculated frequency noise spectra of the QW laser in free-running (burgundy) and in ECL (black) operation.
}
\label{Fig:4}
\end{figure*}

The 35 dB improved feedback insensitivity in the wavelength-detuned QD laser is mainly attributed to the large damping factor \cite{grillot2020physics}. Detailed discussions of the feedback sensitivity are available in the Supplementary Information. Despite the fact that a near-zero LEF is beneficial for suppressing the coherence collapse, it is not always favorable to the ECL since it suppresses the regime II as well \cite{dong2021dynamicdfb}. On the other hand, a large LEF is crucial to realize a significant linewidth reduction \cite{galiev2020optimization}. To suppress the coherence collapse while achieving a large linewidth narrowing, we propose increasing the optical mismatch between the optical gain peak and the Bragg wavelength in the QD laser. In this study, we employ a well-designed DFB grating to enable the laser to operate at 1323 nm, which is 12 nm away from the gain peak at room temperature (Supplementary Information). As a result, the LEF of the lasing mode at the threshold is increased to 2.6, comparable to that of the QW lasers. The consequent increase in the internal loss results in an increase in the threshold current to 60 mA (Fig. \ref{Fig:4}c). Despite the increased LEF, the strong damping of the QD laser can still suppress the coherence collapse. The theoretical analyses of the difference in the feedback sensitivity between the QD and the QW laser are available in Supplementary Information. The simulation results are summarized in Figs. \ref{Fig:3}c and f, which are in good agreement with the experiments. Here, we have validated through the theory that a design of the optical mismatch in a QD DFB laser enables a linewidth reduction regime while maintaining laser coherence. Moreover, the strong damping of the QD laser is advantageous for suppressing the relaxation oscillation frequency, resulting in lower relative intensity noise (RIN). Figure \ref{Fig:3}b depicts the RIN spectra of the free-running QD (burgundy) and the QW (navy) laser at an output power of 0 dBm. The RIN of the QD laser is lower than that of the QW counterpart in the frequency range from 10 MHz to 10 GHz, which is promising for LIDAR applications.

%%%%%%%%%%%%%%%%%%%%%%%%%%%%%%%%%%%%%%%%%%%%%%%%%%%%%%%%%%%%%%%%%%%%%%
%%%%%%%%%%%%%%%%%%%%%%%%%%%%%%%%%%%%%%%%%%%%%%%%%%%%%%%%%%%%%%%%%%%%%%

\medskip
\noindent \textbf{Turnkey locking without power penalty}

The severe coherence collapse in the conventional QW laser can be suppressed through the use of a high-Q external cavity that offers frequency-selective feedback (Fig. \ref{Fig:4}b). This approach provides a narrow frequency filter that only allows the coherent photons to go back to the laser cavity, thus suppressing the coherence collapse of the laser and enabling the QW laser to achieve large linewidth narrowing through the strong backscattering of the microresonator. Nevertheless, this approach requires a high-Q resonator, which can be a challenge to fabricate. Any degradation in the external cavity Q-factor due to fabrication issues can lead to a worsening of laser linewidth locking or even a failure of the SIL. Moreover, anchoring the laser frequency to one of the external cavity resonances demands careful setpoints for laser linewidth locking; the laser also suffers from a power penalty due to the limitation of the external cavity Q-factor. On the contrary, a low-Q external cavity provides a turnkey option for laser linewidth reduction without causing any power penalty (Fig. \ref{Fig:4}a). Nevertheless, the laser can suffer from coherence collapse under strong feedback without a frequency filter.

It used to be a challenge to reduce the QW laser linewidth to hertz level with a low-Q external cavity, due to the weak coherent feedback that it offers. In general, the coherence of QW external-cavity lasers is more than one order of magnitude inferior to the high-Q SIL lasers \cite{tran2019tutorial,morton2021integrated,maier2023sub}. Such a performance gap is attributed to the coherence collapse of the QW laser under strong external feedback. Using the ECL approach by applying the maximum coherent feedback strength of -39 dB, the QW laser linewidth is reduced to 111 Hz, which is 34.9 dB lower than its free-running linewidth (Fig. \ref{Fig:4}g). On the contrary, the chaos-free QD laser allows for bridging the gap between ECL and SIL approaches. In this study, the low-Q external cavity is developed by using optical fiber and a fiber-based reflector without frequency filtering. With both lasers having their output powers fixed at 0 dBm for a fair comparison, a 37.2 dB suppression of the white noise level and a 16 Hz Lorentizan linewidth have been achieved by a QD laser under on-chip resonant feedback strength of -9.6 dB (Fig. \ref{Fig:4}d). The frequency noise reduction is greater than 30 dB across the entire offset frequency range from 1 kHz to 5 MHz. Such a record QD laser linewidth, requiring no sophisticated tuning and locking, is on par with the self-injection-locked QW lasers using a high-Q resonator. The turnkey locking of the QD laser exhibited long-term stability, in the presence of a repeatable narrow linewidth holding at least 40 seconds (Fig. \ref{Fig:3}e.) Moreover, in both experiment and theory, we demonstrate that the laser source does not suffer from a power penalty by using the ECL approach. On the contrary, the laser output power gets amplified in the locking operation owing to the additional coherent photons given by the reflected field for lasing (Fig. \ref{Fig:4}c).

The remarkable ECL performance of the QD laser is twofold: First, the QD laser allows for generating a narrower free-running linewidth than the QW laser owing to its lower LEF. The free-running linewidth of the wavelength-detuned QD laser is 84 kHz, which is four times narrower than its QW counterpart. Second, the chaos-free QD laser does not require a frequency filter to suppress the coherence collapse, thus it can take advantage of the strong coherent feedback offered by a low-Q external cavity. As a result, both the linewidth and the linewidth reduction ratio are improved by utilizing the QD laser. Theoretical analyses of the frequency noise reduction are available in the Supplementary Information. The calculated frequency noise of the QD and the QW laser are shown in Figs. \ref{Fig:4}e and h, respectively, agree with the experiments. Last but not least, it should be noted that the QD laser is also compatible with the high-Q SIL approach and its full potential has yet to be realized. In this work, the 16 Hz intrinsic linewidth is obtained by using an external cavity whose Q-factor is inferior to the Q-factor of the laser cavity ($Q_e / Q_l$ = 0.2 in theory, the blue curve in Fig. \ref{Fig:4}f). Our theory indicates that the laser linewidth can be further reduced to hertz level by increasing the Q-factor of the external cavity beyond that of the laser cavity ($Q_e / Q_l$ = 3.8 in theory, the black curve in Fig. \ref{Fig:4}f).

\begin{table}[!t]
\begin{center}
\caption{Performance comparison of narrow-linewidth lasers using high-Q self-injection locking and low-Q external cavity locking approaches.}
\begin{tabular}{|p{0.2\linewidth}|p{0.24\linewidth}|p{0.15\linewidth}|p{0.2\linewidth}|p{0.1\linewidth}|}
%%{|c|c|c|c|c|}
\hline
Operation principle & Configuration (G/EC)$^{*}$ & Q (M) & Linewidth (Hz) & Ref. No. \\
\hline
ECL$^{\dagger}$ & Monolithic QD-Si/optical fiber & N/A & 16 & This work \\
\hline
SIL$^{\ddagger}$ & Hybrid III-V/Si$_3$N$_4$ & 164 & 0.04 & \cite{li2021reaching} \\
\hline
SIL & Hybrid III-V/Si$_3$N$_4$ & 260 & 1.2 & \cite{Jin2021hertz} \\
\hline
SIL & Heterogeneous III-V/Si$_3$N$_4$ & 50 & 5 & \cite{xiang2023three} \\
\hline
SIL & Heterogeneous III-V/Si$_3$N$_4$ & 7 & 25 & \cite{xiang2021laser} \\
\hline
SIL & Hybrid III-V/Si$_3$N$_4$ & >15 & 25 & \cite{lihachev2022low} \\
\hline
SIL & Heterogeneous III-V/Si$_3$N$_4$ & 10 & 92.4 & \cite{zhang2023photonic} \\
\hline
SIL & Hybrid III-V/Si$_3$N$_4$-LiNbO$_3$ & 1.9 & 3,140 & \cite{snigirev2023ultrafast} \\
\hline
ECL & Hybrid III-V/Si$_3$N$_4$ & N/A & 40 & \cite{fan2020hybrid} \\
\hline
ECL & Heterogeneous III-V/Si$_3$N$_4$ & N/A & 140 & \cite{tran2019tutorial} \\
\hline
ECL & Heterogeneous III-V/Si$_3$N$_4$ & N/A & 400 & \cite{Xiang2021high} \\
\hline
ECL & Monolithic III-V & N/A & 50,000 & \cite{larson2015narrow} \\
\hline
\end{tabular}\\
$^{*}$ G, gain medium; EC, external cavity.
$^{\dagger}$ ECL, external-cavity locking.
$^{\ddagger}$ SIL, self-injection locking.
\end{center}
\label{tab:linewidth}
\end{table}

\medskip
\noindent \textbf{Discussion and outlook} 

This work demonstrates a high-performance QD laser on Si that achieves a Lorentzian linewidth of 16 Hz under external cavity locking. The quantized QD laser opens up more opportunities for large-scale PICs owing to its compatibility with the mature CMOS foundry. Compared to the QD lasers grown on the native substrate, directly growing the QD lasers on Si improves not only the scalability using the commercial 300 mm Si wafers but also the laser performance. For instance, the epitaxial QD laser on Si enabled a record LEF as low as 0.13 \cite{duan2018semiconductor}, whereas the LEF reported in the native QD laser was in general above 0.7 \cite{duan2018narrow}. Such an improvement results from the optimization of the inhomogeneous broadening and the defects. The shortening of Shockley-Read Hall (SRH) recombination time also contributes to the near-zero LEF in the epitaxial QD laser on Si \cite{zhao2021effect}. Superior to conventional QW heterostructure, the advanced QD enables monolithic integration of laser sources, amplifiers, and detectors on Si, which is a low-cost solution for Si PICs. After optimizing the fabrication, the QD DFB laser performance has been improved. This includes the decrease in the threshold current density to 400 A/cm$^{2}$, the increase in the SMSR to 59 dB, the reduction in the intrinsic linewidth to 41 kHz, and the improvement in the yield \cite{wan20201}. Previous studies have shown that the near-zero LEF offered by the QD laser allows for isolator-free photonic integration \cite{duan20191}. Here, we reveal that the LEF of the QD laser can be practically engineered by the wavelength detuning between the optical gain peak and the Bragg wavelength. As such, an increase in the LEF through a wavelength-detuned DFB grating contributes to a large linewidth reduction while maintaining the laser's coherence. By leveraging the chaos-free operation of the QD laser, its linewidth can be locked to a low-Q external cavity without the need for a frequency filter. Table I\ref{tab:linewidth} compares the current state-of-the-art narrow-linewidth lasers using either the high-Q SIL approach or the low-Q ECL approach. While an ultra-high-Q resonator using the low-loss SiN platform enables hertz level and sub-hertz level laser linewidth \cite{Jin2021hertz,li2021reaching,xiang2023three}, fabrication is challenging as any degradation in the Q-factor would result in a broader laser linewidth \cite{xiang2021laser,lihachev2022low,zhang2023photonic}. In addition, achieving ultra-high-Q resonators in many material platforms, such as lithium niobate (LN) \cite{Zhang:17}, poses significant challenges. As a result, the locked linewidth of SIL lasers using LN resonators is on the order of kilohertz \cite{snigirev2023ultrafast}, which remains to be improved for coherent LIDAR applications. Our work, however, demonstrates that the locked QD laser linewidth is on par with or even better than most high-Q SIL QW lasers. This finding underscores the potential of QD lasers in applications requiring precise coherence, particularly where achieving ultra-high-Q resonators is challenging. In addition, our approach provides additional benefits including turnkey and power-maintaining operations. On the other hand, employing a low-Q external cavity can potentially enhance operation stability due to its broader reflection bandwidth, which is more accommodating of laser wavelength shifts. Along with the high thermal stability of epitaxial QD lasers on Si - capable of continuous-wave operation above 150  $^{\circ}$C \cite{wang2023inas} - this approach is beneficial to the robustness of ECL in high-temperature operation.

Further improving the laser coherence can be effectively achieved with an on-chip external cavity solution. Using a fiber-based external cavity, the feedback strength is mainly limited by the laser-to-fiber coupling loss. By leveraging the laser and the external cavity on the same chip, this coupling loss can be largely reduced. 3D photonic integration can potentially leverage the ultra-low-loss SiN waveguide, which is advantageous for maximizing the on-chip coherent feedback strength and reducing laser linewidth \cite{xiang2023three}.

Another approach to improve the SIL is to apply a frequency filter by using a high-Q resonator. The QD laser is also compatible with the CMOS-ready SiN and silica microresonators that offer 1 billion Q-factor \cite{wu2020greater,liu2022ultralow}. We theoretically demonstrate that sub-hertz level linewidth can be achieved by using the QD laser and a high-Q resonator \cite{chow2022analysis,alkhazraji2023linewidth}.Additionally, spiral microresonators with large modal volumes can be employed to reduce the low-offset frequency noise caused by thermodynamic fluctuations \cite{li2021reaching}. Depending on the design of the external cavity, the high-coherence QD laser can scale up to 300 mm Si wafers using either monolithic or heterogeneous photonic integration technologies. This would lead to low-cost, large-volume, and scalable production. The potential for monolithic integration of III-V lasers and passive waveguides onto a single chip using foundry-based technologies could further advance the development of QD lasers and their integration into practical photonic applications.

\medskip
\begin{footnotesize}

\noindent \textbf{Data Availability}: 
All data generated or analyzed during this study are available within the paper and its Supplementary Information. 
Further source data will be made available online once published.

\noindent \textbf{Code Availability}: 
The analysis codes will be made available on reasonable request.

\noindent \textbf{Acknowledgments}: 
This work is supported by the Defense Advanced Research Projects Agency (DARPA) MTO GRYPHON (HR0011-22-2-0009) and LUMOS (HR0011-20-2-0044) programs. We thank Emad Alkhazraji for fruitful discussions and figure polish.

\noindent \textbf{Author contribution}: 
B.D. and Y.W. led the QD laser design and device characterization. 
Y.W. designed and fabricated the QD lasers with assistance from C.S. and R.K.
B.D. characterized and gathered the experimental data from the device, with contributions from R.K..
W.W.C. provided theoretical calculations and analysis on the locking dynamics and feedback sensitivity, with contributions from A.P. and H.W..
B.D. wrote the manuscript with inputs from Y.W., C.S. and A.P.. 
All authors commented on and edited the manuscript.
J.E.B supervised the project.

\noindent \textbf{Competing Interests}: 
J.E.B.is a cofounder and shareholder of Nexus Photonics and Quintessent, startups in silicon photonics.

\end{footnotesize}

\bibliographystyle{apsrev4-2}
\bibliography{bibliography}

% \pagebreak

\bigskip

\medskip
\noindent \textbf{Methods} 

\medskip
\noindent \textbf{Free-running laser characterizations}

The optical spectra are measured using a high-resolution Optical Spectrum Analyzer (OSA) (Yokogawa AQ6370C). The laser frequency noise and resultant fundamental linewidth are taken from a commercial laser phase noise analyzer (PNA) (OEwaves OE4000) that internally performs averaging over the measured phase noise. As automated equipment, this PNA offers a measurement bandwidth of 150 MHz. An average of 50 was applied in the range of 1 kHz to 150 MHz to measure the frequency noise. In the measurements of optical spectra, the QD laser is driven by a programmable current source (Keithley 2400). It is then driven by a low-noise laser current source (ILX Lightwave LDX-3620) to ensure stable and low-noise operation in the measurement of frequency noise. The laser relative intensity noise (RIN) is measured by a 20 GHz photodetector (Agilent 11982A). The converted AC signal filtered by a bias-tee is preamplified by a 30 dB RF amplifier (RF-Lambda RLNA00G18GA) before it is sent to an Electrical Spectrum Analyzer (ESA) (Rohde \& Schwarz FSU50). The resolution bandwidth (RBW) used in the RIN measurement is 1 MHz. The RIN bandwidth is limited by the RF amplifier, which is 18 GHz.

\noindent \textbf{Laser external cavity locking}

Figure \ref{Fig:3}a depicts the experimental configurations for laser external cavity locking. The coupled laser emission is sent to a 90/10 fiber beam splitter, after which 90\% of the coupled power will be used for external optical feedback. The feedback loop consists of a polarization controller (PC), a variable optical attenuator (VOA), and a back-reflector (EXFO OSICS BKR). The external cavity frequency is 6 MHz, which corresponds to 16.6 m-long optical fiber. The remaining 10\% of the laser output is utilized for the characterizations of feedback sensitivity and laser linewidth. After passing through an optical isolator, it is transferred either to a phase noise analyzer (PNA) (OEWaves OE4000) for frequency noise characterization or a delayed self-heterodyne setup to check the electrical spectrum-based laser coherence. 

In this study, the feedback strength is determined by the on-chip reflected power ($P_{refl}$) and the free-space output power ($P_{out}$) through the following relationship: 
\begin{equation}
\eta_{F}=\frac{P_{refl}}{P_{out}}
\label{equ:fb}
\end{equation}

All losses from the external cavity should be considered to calculate the reflected power, thus the feedback strength. After optimizing the setup, the fiber-chip coupling loss is -2.4 dB (round-trip coupling loss is -4.8 dB), the total losses from the 90\% beam splitter, the insertion loss of the VOA, the polarization controller, and the fiber is -4.81 dB. The maximum on-chip feedback strength that we can achieve is -9.61 dB.

In the external cavity locking operation, the laser is driven with low-noise laser current sources (ILX Lightwave LDX-3620) to ensure stable and low-noise operation. Detection of the locking state is assured by the reduction in the linewidth of the self-heterodyne beat. The self-heterodyne interferometer setup consists basically of a Mach-Zehnder interferometer (made from two 3 dB couplers) with a PC and a short delay line in one of its arms and a fiber-coupled acoustic-optic modulator (Gooch \& Housego 27 MHz) in the other arm, as shown in Fig.~\ref{Fig:3}a. The beat frequency from the self-heterodyne interferometer is detected using a photodetector (Newport 1811) before it is sent to an Electrical Spectrum Analyzer (ESA) (Rohde \& Schwarz FSWP). 

To investigate the turnkey locking stability, we periodically vary the on-chip feedback strength between -61.2 dB and -9.61 dB. The periodicity of the feedback strength calibration is 80 seconds. The RBW of the ESA was 10 Hz in this measurement.

\medskip

\pagebreak

\end{document}